\documentclass[11pt,a4paper]{article}
\usepackage{jcappub}
\notoc
\usepackage[utf8]{inputenc}
\usepackage[T1]{fontenc}
\usepackage{lipsum}
\usepackage{graphicx}
\usepackage{amsmath}
\usepackage[normalem]{ulem}
\usepackage{amsfonts}
\usepackage{amssymb}
\usepackage{calrsfs}
\usepackage[mathscr]{euscript}
\usepackage{multirow}
\usepackage{natbib}
\usepackage{tikz}
\usepackage[colorlinks=true,linktocpage=true,linkcolor=blue,citecolor=blue]{hyperref}
\usepackage{booktabs}

\definecolor{mypink}{RGB}{219, 48, 222}
\definecolor{brown}{RGB}{200,150, 50}

\newcommand{\tr}{\rm{Tr}}

\title{Hadron-quark phase transition: the QCD phase diagram and stellar conversion}

\author[a]{Clebson A. Graeff,}
\author[b]{Marcelo D. Alloy,}
\author[c]{Kauan D. Marquez,}
\author[d]{Constança Provid\^encia}
\author[c]{and D\'ebora P. Menezes}

\affiliation[a]{Universidade Tecnol\'ogica Federal do Paran\'a, campus
  Pato Branco, \\ Via do Conhecimento, Km 1 CEP 85503-390 Pato Branco
  -- PR, Brazil}
\affiliation[b]{Departamento de Ci\^encias Exatas e Educa\c{c}\~ao, Universidade Federal
de Santa Catarina \\ Blumenau, SC, CEP 89.065-300, Brazil}
\affiliation[c]{Departamento de F\'isica, Universidade Federal de Santa Catarina \\
Florian\'opolis, SC, CP 476, CEP 88.040-900, Brazil}
\affiliation[d]{Departamento de F\'isica, Universidade de Coimbra, 3004-516 Coimbra, Portugal}

\emailAdd{cgraeff@utfpr.edu.br}
\abstract{
Different extensions of the Nambu-Jona-Lasinio model, known to satisfy
expected QCD chiral symmetry aspects, are used to investigate a possible 
hadron-quark phase transition at zero temperature and to build the
corresponding binodal sections. We have shown that the transition
point is very sensitive to the model parameters and that both pressure
and chemical potential increase drastically with the increase of the
vector interaction strength in the quark sector. 
Within the same framework, the possibility of quark  and  hybrid star
formation is analyzed.
The same conclusions drawn before with respect to the 
coexistence pressure and chemical potentials
are reinforced. We conclude that even if a transition from a metastable hadronic star  to a quark
star is thermodinamically possible, it is either energetically
forbidden or gives rise to a blackhole. Nevertheless, conversions from
metastable to hybrid stars are possible, but the  mass difference between
both compact objects is very small, never larger than 0.2 $M_\odot$.
}


\arxivnumber{1806.04170}

\begin{document}
%%%%%%%%%%%%%%%%
%%%%%%%%%%%%%%%%

\maketitle

%%%%%%%%%%%%%%%%%%%%%%
\section{Introduction}
%%%%%%%%%%%%%%%%%%%%%%

The complete understanding of the quantum chromodynamics (QCD) phase diagram represents a challenge in both theoretical and experimental physics. While many of the features it aims to describe can be tested in heavy-ion collision experiments, other aspects related to matter under extreme conditions can only be inferred from results of lattice QCD (LQCD) \cite{lqcd} or from observational results of astrophysical objects.

Effective models remain a good source of information about regions of the QCD phase diagram inaccessible by terrestrial experiments or by LQCD methods, providing qualitative results and theoretical insights. The present work intends to help in advancing
our knowledge towards some of the regions of the QCD phase diagram through this strategy. When this approach is applied to the study of the transition of hadronic matter to the deconfined quark matter, it is suggested that the QCD phase diagram shows a first order phase transition \cite{1995NuPhA.590..367K}.

From the LQCD perspective, the hadron-quark transition at zero
chemical potential is believed to be a crossover. However the whole
diagram is not expected to be covered  within this approach in a near future,
due to some numerical difficulties as the {\it sign problem} and to the
 huge computational cost, so that only the region close to
zero chemical potential has been assessed so far. { Effective models,
such as the NJL  \cite{NJL} and PNJL \cite{pnjl},
have, therefore, been used to study the phase diagram at high chemical
potential. Within these models, it is expected
that the crossover at zero chemical potential and high temperature
goes into a first order phase transition at high chemical potential and
low temperature \cite{njl-cep,pnjl-cep}. }

This idea is reinforced by experimental results, such as the Beam Energy Scan (BES-I,
BES-II) programs at RHIC that have provided 
data signaling to a first order phase transition and pointed out
the possible existence and location of the critical end point
\cite{BES}. Future
experiments to take place at the FAIR facility at GSI \cite{FAIR},
NICA at the JINR \cite{NICA} and the NA61/SHINE program at SPS (CERN)
\cite{NA61} will also contribute to shed some light into these still
unknown aspects. Also, the improvement of the observational results of
compact objects are expected from the new paradigm of multimessenger
astronomy established by the gravitational waves detection of compact
star mergers \cite{ligo}, and from x-ray telescopes, such
as NICER \cite{NICER} and, in the future, Athena \cite{ATHENA}, which
will provide better constraints to the effective model
parameterizations and thus contribute to the understanding of the low
temperature and high density regime of the QCD phase diagram. A
  comprehensive historical perspective  on the nuclear astrophysics
  aspects in the new era of multimessenger astronomy can be found in \cite{jorge_hadrons}.

It is argued that the phase transition in QCD can take place in two different steps at low temperatures, first by the (partial) restoration of the chiral symmetry where chromodynamic matter is still confined, giving rise to the so-called quarkyonic phase \cite{quarkyonic}, and only then by the deconfination phase transition. 

Considerations on the phase transition at zero temperature have
already been done in many works \cite{muller1997, ditoro2006,
Cavagnoli2011, Tsue2010, Lee2013, marquez17}, but we do believe the
formalism we present next is more adequate, because the effective
models employed here exhibit chiral symmetry in both hadronic and
quark phases, which is demanded to take seriously the appearance of
the quarkyonic phase. The models to be used here are all included in
the Nambu--Jona-Lasinio (NJL) model framework \cite{NJL} in order to
naturally describe the chiral characteristics of QCD matter. 

The two EoS model has been applied to the study of the hadron-quark
phase transition in several studies,
\cite{DiToro2009,marquez17,Cavagnoli2011,Shao2011,Shao2012,DiToro2016}. In
particular, in \cite{Shao2012}  the effect of the vector contribution
in the quark-matter description was considered. In this studies the
authors have discussed the hadron-quark phase transition taking for
nuclear matter a  RMF model and for quark matter the PNJL model
including vector terms. A review of the effect of the symmetry energy
on the hadron-quark phase transition was presented in
\cite{DiToro2016} where the possible signatures  of the
hadron-quark phase transition in the range
of the NICA program are also discussed. It is the aim of the present
work to further study this phase transition taking the two EoS model
but using, for the first time,  the hadron and the quark phase
chiral symmetric models for both EoS. This will be an exploratory study and,
therefore, we will restrict ourselves to zero temperature.

 In \cite{Cavagnoli2011} and \cite{marquez17}, the hadron-quark phase transition was
investigated with the help of two different models, namely, the
non-linear Walecka model (NLWM) for the hadronic phase and the MIT bag
model for the quark phase. A formalism we understand as a more
adequate one was used in \cite{Providencia2003,Providencia2003a,Tsue2010} at zero temperature and by \cite{Lee2013} for finite
temperatures, all considering NJL-type models for
the two phases. To describe the hadron phase, the standard NJL model with
vector interaction is extended to include a scalar-vector channel in order to render the model capable of saturation at low densities \cite{Koch1987}.
In the present work, we revisit the approach of~\cite{Tsue2010}
and~\cite{Lee2013}, but applying an extended NJL model for the hadron 
phase that includes  additional channels to achieve a better 
description of important nuclear bulk properties \cite{Pais2016}.
A similar extension of the NJL model for hadronic matter has been developed
\cite{Nguyen1,Nguyen2} with a different choice of interaction channels.
Recently, this version was also applied to investigate the
hadron-quark phase transition \cite{Nguyen3}, but the quark phase was
still described by the MIT bag model. 
Albeit the authors of the afore mentioned papers also refer to the hadronic extension as eNJL, here we
name PPM NJL the more complete extension developed in \cite{Pais2016}, to avoid confusion with the
versions proposed in \cite{Koch1987} and \cite{Nguyen1,Nguyen2}.

Hence, in the present work, we describe the hadronic matter with the
PPM NJL model and the quark matter with the NJL in its SU(2) version in order to check for which
parameters of these two models the phase transition is possible, 
considering both symmetric and asymmetric systems. Whenever possible,
the binodal sections are obtained. We  include in the quark
  model a vector contribution that has proved to make the quark EoS
  stiffer and may have important consequences on the structure of
  hybrid or pure quark compact stars
\cite{Hanauske2001,Klahn2007,Bonanno2011,PhysRevD.94.094001}.
 In particular, the inclusion of this term gives rise  to larger star
 masses although with smaller quark cores in the case of hybrid stars.

In the sequel, as an application of  our two phase model and to
  compare our calculations with already existing results in the
  literature  we impose $\beta$-equilibrium and charge neutrality
conditions to obtain equations of state (EoS) for both phases in order
to investigate the possibility of a hadron-quark phase transition to
occur in the interior of compact stars. For this study, we 
consider the SU(3) version of the NJL model for the quark phase so
that the strangeness demanded by the Bodmer-Witten conjecture for the
stability of quark matter \cite{bodmer,witten} is considered,
  although we are aware that this model does not produce absolutely
  stable matter at zero temperature. We discuss this in more
  detail during the presentation of the results.
These EoS are then applied to describe respectively hadronic, hybrid and quark stars to check
when the former is metastable to decay into the two latter and
 the possibility of a hadron-quark conversion inside these objects is checked.
 
This paper is organized as follows: in Section II we describe the basic
formalism necessary to the understanding of the NJL model in all
versions we need. In Section III we discuss how to
obtain the binodal points at zero temperature, display and comment our
results. In Section IV, we investigate if a hadron-quark phase
transition can take place inside a compact star, the
existence of metastable hadronic stars and the possible conversion to stable quark
stars. Finally, in the last section, we make some final remarks and
discuss the continuation of the present work.

%%%%%%%%%%%%%%%%%%%
\section{Formalism} 
%%%%%%%%%%%%%%%%%%%

In the following we present the basic equations underlying the NJL
models in three different versions: the usual SU(2) and SU(3) versions
that describe quark matter and the PPM version of the NJL model that
describes hadronic matter. 

%%%%%%%%%%%%%%%%%%%%%%%%%%%%%%%%%%%%%
\subsection{Quark Matter - NJL SU(2)}\label{NJLSU2}
%%%%%%%%%%%%%%%%%%%%%%%%%%%%%%%%%%%%%

The quark phase is described by a SU(2) NJL model Lagrangian including a vector term, given by \cite{Buballa2005}
\begin{equation}\label{Eq:LagNJL-SU2-Bub}
    \mathscr{L} = \bar{\psi}(i\gamma^\mu\partial_\mu - \hat m_0)\psi + G_s[(\bar{\psi}\psi)^2 + (\bar{\psi}i\gamma_5\vec{\tau}\psi)^2] - G_v(\bar{\psi}\gamma^\mu \psi)^2.
\end{equation}
Here $\psi$ represents the quark field, $\hat m_0$  the quark bare mass, and
$G_s$ and $G_v$ are coupling constants that are fitted by the
pion mass $m_\pi = 135.0~\rm{MeV}$ and its decay constant $f_\pi =
92.4~\rm{MeV}$. As a non-renormalizable theory, a momentum cutoff
$\Lambda$ must be employed in the momenta, which acts as a new free parameter of the model.
In Table \ref{Tab:Parametros_NJL}, four possible parameter sets usually
considered in the literature are given. Notice that the $G_v$
parameter can be arbitrarily chosen, allowing to write 
$G_v=xG_s$, where $x$ is a free parameter varying in the range $0\leq x\leq1$ \cite{PhysRevD.64.043005}. Furthermore in text the parameterization choice is written as, e.g., PCP-0.1, which reads as the PCP parameter set taken with x=0.1.

From the Lagrangian $\mathscr{L}$, one can obtain the thermodynamic potential per volume $V$ at temperature $T$ through the Hamiltonian density $\mathscr{H}$, which leads to
\begin{equation}
	\Omega(T, \mu) = -\frac{T}{V} \ln \tr \exp\left[-\frac{1}{T}\int d^3x(\mathscr{H} - \mu\psi^\dagger\psi)\right],
\end{equation}
where $\tr$ stands for the trace over all states of the system, resulting in \cite{Buballa1996,Buballa2005}
\begin{equation}\label{Eq:Pot_Termo_Temp_zero}
	\Omega(\mu; M, \widetilde{\mu}) = \sum_{i = u,d} \Omega_{M_i}(T,\mu_f)  + G_s(\phi_u + \phi_d)^2  - G_v(\rho_u + \rho_d)^2,
\end{equation}
with, at the zero temperature limit in the mean-field approximation,
\begin{equation}
	\Omega_{M_i} = -2 N_c \int \frac{d^3p}{(2\pi)^3} \left[E_p + (\widetilde{\mu}_i - E_p) \theta(p_{F_i} - p)\right],
\end{equation}
$\mu$ stands for the chemical potential and $M_i$ are the quarks constituent masses. Here $N_c$ stands for the number of colors and $E_p^i = \sqrt{p^2_f + M_i^2}$. The Fermi momentum of the quarks is represented by $p_{F_i}$ and $\theta(p_{F_i} - p)$ stands for the step function.

The renormalized chemical potential $\widetilde{\mu}_i$ and the constituent mass $M_i$ are respectively obtained by requiring $\partial \Omega / \partial \widetilde{\mu}_i = 0$ and $\partial \Omega / \partial M_i = 0$, resulting in
\begin{align}
	\widetilde{\mu}_i &= \mu_i - 2 G_v \rho, \qquad i = {u, d}, \\
	M_i &= m - 2 G_s\phi,  \qquad i = {u, d},
\end{align}
with $\rho=\rho_u + \rho_d,$ $\phi=\phi_u + \phi_d$, and
\begin{align}
	\rho_i &= \frac{N_c}{3\pi^2} p_{F_i}^3 \label{rhobar}\\
	\phi_i &= \langle \bar\psi\psi\rangle = - 2 N_c \int_0^\Lambda\frac{d^3p}{(2\pi)^3} \frac{M_i}{E_p^i}\left[1 - \theta(p_{F_i} - p)\right], \label{phii} \\
	\widetilde{\mu}_i &= \sqrt{p_{F_i}^2 + M_i^2}.
\end{align}

A constant term in the potential has no physical meaning, consequently such term may be chosen so that the thermodynamic potential is zero at the value $M = M_{\rm{vac}}$ which minimizes $\Omega$ at $T = \mu = 0$. This process may be represented by 
\begin{equation}
\widetilde\Omega(\mu; M, \widetilde{\mu}) = \Omega(\mu; M, \widetilde{\mu}) - \Omega(0,0; M_{\rm{vac}}, 0),
\end{equation}
so that the pressure $P$ and the energy density $\varepsilon$, the quantities we are interested in, are obtained through
\begin{equation}
P = -\widetilde\Omega(\mu; M, \widetilde\mu), \label{Exp_pressao_T}
\end{equation}
and
\begin{equation}
 \varepsilon = -P + \sum_{i=u,d} \mu_i \rho_i, \label{Exp_energia_T}
\end{equation}
respectively resulting in
\begin{equation}
    P = 2N_c \sum_{i=u,d}\left[ \int_{p_{F_i}}^\Lambda \frac{d^3p}{(2\pi)^3} \, E_{p}^i + \frac{\widetilde\mu_i p_{F_i}}{6\pi^2} \right] - G_s(\phi_u + \phi_d)^2 + G_v(\rho_u + \rho_d)^2  + \Omega(0, 0; M_{\rm{vac}}, 0)
\end{equation}
and
\begin{equation}
    \varepsilon = 2N_c \sum_{i=u,d}\left[ - \int_{p_{F_i}}^\Lambda \frac{d^3p}{(2\pi)^3} \, E_{p}^i + (\mu_i - \widetilde\mu_i) \frac{p_{F_i}}{6\pi^2} \right] + G_s(\phi_u + \phi_d)^2 - G_v(\rho_u + \rho_d)^2 - \Omega(0, 0; M_{\rm{vac}}, 0).
\end{equation}
The equations are then solved self-consistently for each value of $\rho_i$ or $\mu_i$, noticing that one has ${p_{F_i} = \sqrt{\widetilde\mu_i^2 - M_i^2}}$ for ${\widetilde\mu_i^2 \geq M_i^2}$ in this case. All the parameters are presented in Table \ref{Tab:Parametros_NJL}.

\begin{table}[!t]
\centering
\caption{Parameter sets for the SU(2) NJL Lagrangian density~\eqref{Eq:LagNJL-SU2-Bub} \cite{Buballa1996, Buballa2005,PhysRevD.94.094001}. \label{Tab:Parametros_NJL}}
\begin{tabular}{lccccc}
\toprule
Model &  $\Lambda$ & $G_s$ ($\rm{fm}^2$) & $G_v$ ($\rm{fm}^2$) & $m_0$ (MeV) & $M$ (MeV) \\
\midrule
Buballa-1 & 650 & 0.19721 & -- & 0 & 313 \\
Buballa-2 & 600 & 0.26498 & -- & 0 & 400 \\
BuballaR-2 & 587.9 & 0.27449 & $\propto G_s$ & 5.6 & 400 \\
PCP & 648 & 0.19565 & $\propto G_s$ & 5.1 & 312.6 \\
\bottomrule
\end{tabular}
\end{table}

%%%%%%%%%%%%%%%%%%%%%%%%%%%%%%%%%%%%%
\subsection{Quark Matter - NJL SU(3)}\label{NJLSU3}
%%%%%%%%%%%%%%%%%%%%%%%%%%%%%%%%%%%%%

\begin{table*}[!t]
\centering
\caption{Parameter sets for the SU(3) NJL Lagrangian
  density~\eqref{Eq:LagNJL-SU3} \cite{HK94,PhysRevD.94.094001} 
\label{Tab:Parametros_NJL_SU3}. }
\begin{tabular}{lcccccccc}
\toprule
Model &  $\Lambda$ & $G_s \Lambda^2$ & $K \Lambda^5$ & $G_v$ 
& $m_{u,d}$ (MeV) & $m_s$ (MeV) & $M_{u,d}$ (MeV)  & $M_s$ (MeV) \\
\midrule
HK & 631.4   & 1.835 & 9.29 &  $x G_s$ & 5.5 & 135.7 & 335.5 & 528 \\
PCP & 630.0 & 1.781 & 9.29 &  $x G_s$ & 5.5 & 135.7 & 312.2 & 508 \\
\bottomrule
\end{tabular}
\end{table*}

\begin{table*}[!t]
\centering
\caption{Parameter sets for the PPM NJL Lagrangian density \eqref{Eq:Lagrangiana_eNLJ_Pais} \cite{Pais2016}. ($G_s$, $G_v$, and $G_\rho$ values are in $\rm{fm}^2$; $G_{sv}$, $G_{v\rho}$, and $G_{s\rho}$ in $\rm{fm}^8$.) \label{Tab:Parametros_eNJL}}
\begin{tabular}{lcccccccc}
\toprule
Model & $G_s$ & $G_v$ & $G_{sv}$ & $G_\rho$ & $G_{v\rho}$ & $G_{s\rho}$ & $\Lambda$ (MeV) & $m$ (MeV) \\
%Model & $G_s$ ($\rm{fm}^2$) & $G_v$ ($\rm{fm}^2$) & $G_{sv}$ ($\rm{fm}^8$) & $G_\rho$ ($\rm{fm}^2$) & $G_{v\rho}$ ($\rm{fm}^8$) & $G_{s\rho}$ ($\rm{fm}^8$) & $\Lambda$ (MeV) & $m$ (MeV) \\
\midrule
eNJL3$\sigma\rho$1 & 1.93 & 3.0 & -1.8 & 0.0269 & 0 & 0.5 & 534.815 & 0 \\
eNJL2m$\sigma\rho$1 & 1.078 & 1.955 & -2.74 & -0.1114 & 0 & 1 & 502.466 & 500 \\
\bottomrule
\end{tabular}
\end{table*}

Dense matter in a quark phase can also be described with the
SU(3) version of NJL model, which incorporates the
$s$-quark, with the repulsive vector interaction. In this case,
the Lagrangian density is given by \cite{njlv,PhysRevD.94.094001}
\begin{equation}
\mathscr{L}=\bar{\psi}\left(i\gamma_\mu\partial^\mu-\widehat{m}_f\right)\psi+
\mathscr{L}_{\rm{sym}}+\mathscr{L}_{\rm{det}}+\mathscr{L}_{\rm{vec}},
\label{Eq:LagNJL-SU3}
\end{equation}
with $\mathscr{L}_{\rm{sym}}$, $\mathscr{L}_{\rm{det}}$ and $\mathscr{L}_{\rm{vec}}$ given by
\begin{eqnarray}
\mathscr{L}_{\rm{sym}}&=&G_s\sum_{a=0}^8[(\bar{\psi}\lambda_a\psi)^2+(\bar{\psi}i\gamma_5\lambda_a\psi)^2],\nonumber\\
\mathscr{L}_{\rm{det}}&=&-K\{\mbox{det}[\bar{\psi}(1+\gamma_5)\psi]+\mbox{det}[\bar{\psi}(1-\gamma_5)\psi]\},\nonumber\\
%\mathscr{L}_{vec}&=&-G_s\sum_{a=0}^8[(\overline{\psi}\gamma_\mu\lambda_a\psi)^2+(\overline{\psi}\gamma_5\gamma_\mu\lambda_a\psi)^2],
\mathscr{L}_{\rm{vec}}&=&-G_v(\bar{\psi}\gamma^\mu\psi)^2,\nonumber
\end{eqnarray}
where $\psi(u,d,s)$ represents the three flavor quark field,
$\widehat{m}_f=\mbox{diag}(m_u,m_d,m_s)$ is the quark current mass matrix,
$\lambda_0=\sqrt{2/3}I$ where $I$ is the U(3) unit matrix, and
$\lambda_a$, with $a=1,\ldots,8$, are the Gell-Mann flavor matrices. 

To obtain effective quark masses $M_i$ we must minimize the thermodynamic potential 
given by
\begin{equation}
    \Omega(\mu; M, \widetilde\mu) = \sum_{i=u,d,s}\Omega_{M_i}(T,\mu_f) + 2G_s (\phi_u^2 + \phi_d^2 + \phi_s^2) - 4K\phi_u \phi_d \phi_s - 2G_v(\rho_u + \rho_d + \rho_s)^2,
\end{equation}
with, at the zero temperature limit in the mean-field approximation,
\begin{equation}
	\Omega_{M_i} = - 2 N_c \int_{p_{F_i}}^\Lambda \frac{d^3p}{(2\pi)^3} \frac{p^2 + m_i M_i}{E_p^i}
\end{equation}
where $\phi_i$ and $\rho_i$ are the same as in the SU(2) case.
The renormalized chemical potential $\widetilde{\mu}_i$ is given by
\begin{equation}
\widetilde{\mu}_i=\mu_i-2G_v\rho, \qquad \rho=\rho_u+\rho_d+\rho_s,
\end{equation}
where $i$ refers to the flavor and $\rho_i$ refers to the respective quark number density. Thus,
minimizing $\Omega$, we obtain in mean-field approach the following gap equations
\begin{equation}
M_i=m_i-4G_s\phi_i+2K\phi_j\phi_k,
\end{equation}
with $(i,j,k)$ being any permutation of $(u,d,s)$.

Then, using (\ref{Exp_pressao_T}) and (\ref{Exp_energia_T}), the pressure and energy density may be written as
\begin{equation}
 \begin{aligned}
    P =&~ 2 N_c \sum_{i=u,d,s}\left[ \int_{p_{F_i}}^{\Lambda} \frac{d^3p}{(2\pi)^3} \frac{p^2 + m_i M_i}{E_p^i} \right] - 2G_s (\phi_u^2 + \phi_d^2 + \phi_s^2) + 2K\phi_u \phi_d \phi_s \\
    &~ + 2G_v(\rho_u + \rho_d + \rho_s)^2 + \Omega(0, 0; M_{\rm{vac}}, 0) 
    \end{aligned}
\end{equation}
\begin{equation}
 \begin{aligned}
    \varepsilon =&~ 2N_c\sum_{i=u,d,s}\left[\mu_i\rho_i -\int_{p_{F_i}}^{\Lambda}\frac{d^3p}{(2\pi)^3}\frac{p^2+m_iM_i}{E_p^i}\right] +2G_s(\phi_u^2 + \phi_d^2 + \phi_s^2)-2K \phi_u \phi_d \phi_s \\
    &~ -2 G_v(\rho_u + \rho_d + \rho_s)^2 - \Omega(0, 0; M_{\rm{vac}}, 0),
        \end{aligned}
\end{equation}
The parameter sets used for the SU(3) NJL are shown in Table
\ref{Tab:Parametros_NJL_SU3}, where again
$G_v=xG_s$.

%%%%%%%%%%%%%%%%%%%%%%%%%%
\subsection{Hadron Matter - PPM NJL}\label{NJLHM} 
%%%%%%%%%%%%%%%%%%%%%%%%%%

Even though the original NJL model is unable to describe the
saturation properties of the nuclear matter, this can be fixed by an
extended version which includes a scalar-vector channel. This gives origin to the so-called extended Nambu--Jona-Lasinio
(eNJL) models \cite{Koch1987}. Many other channels of interaction can  also be
included in the original NJL Lagrangian density, in order to help the description
of other important bulk properties, as discussed ahead. This rather broaden model are known as PPM NJL model, given by \cite{Pais2016}
\begin{equation}\label{Eq:Lagrangiana_eNLJ_Pais}
\begin{split}
	\mathscr{L} =&~ \bar{\psi}(i\gamma^\mu\partial_\mu - \hat m)\psi + G_s[(\bar{\psi}\psi)^2 + (\bar{\psi}i\gamma_5\vec{\tau}\psi)^2]  - G_v(\bar{\psi}\gamma^\mu\psi)^2 \\
	& - G_{sv}[(\bar{\psi}\psi)^2 + (\bar{\psi}i\gamma_5\vec{\tau}\psi)^2](\bar{\psi}\gamma^\mu\psi)^2 - G_\rho[(\bar{\psi}\gamma^\mu\vec{\tau}\psi)^2 + (\bar{\psi}\gamma_5\gamma^\mu\vec{\tau}\psi)^2] \\
	& - G_{v\rho}(\bar{\psi}\gamma^\mu\psi)^2[(\bar{\psi}\gamma^\mu\vec{\tau}\psi)^2 + (\bar{\psi}\gamma_5\gamma^\mu\vec{\tau}\psi)^2] \\
	& - G_{s\rho} [(\bar{\psi}\psi)^2 + (\bar{\psi}i\gamma_5\vec{\tau}\psi)^2] \times [(\bar{\psi}\gamma^\mu\vec{\tau}\psi)^2 + (\bar{\psi}\gamma_5\gamma^\mu\vec{\tau}\psi)^2],
\end{split}
\end{equation}
where $\psi$ represents the nucleon field and the constants $G_i$
stand for the coupling constants for the different channels: the $G_v$
term simulates a chiral-invariant
short-range repulsion between the nucleons, the $G_{sv}$ term
accounts for the density dependence of the scalar coupling, the 
$G_\rho$ term  allows for the description of isospin asymmetric matter, and
the $G_{\omega\rho}$ and $G_{s\rho}$ terms make the density dependence
of the symmetry energy
softer.  
For nuclear matter, the NJL model leads to binding, but the
binding energy per particle does not have a minimum except at a rather
high density where the nucleon mass is small or vanishing. The
introduction of the term in $G_{sv}$ helps to correct this unwanted feature.
As in the quark case, the theory is renormalized via a three momenta cutoff
$\Lambda$. In Table \ref {Tab:Parametros_eNJL}, the parameter sets
used in the present work are given. They were chosen because they were
successful in describing all important nuclear and stellar
constraints, as can be seen in \cite{Pais2016}.

The thermodynamic potential is obtained from~\eqref{Eq:Lagrangiana_eNLJ_Pais} in the same way as for the quark case, and is given by
\begin{equation}\label{Eq:potencial_termodinamico}
	\Omega(\mu) = \varepsilon_{\rm{kin}} + M\phi - \mu_p\rho_p - \mu_n\rho_n - G_s\phi^2 + G_v\rho_B^2+ G_{sv}\phi^2\rho_B^2 + G_\rho\rho_3^2+ G_{v\rho}\rho_B^2\rho_3^2 + G_{s\rho}\phi^2\rho_3^2,
\end{equation}
where the kinetic energy contribution is given by
\begin{equation}
	\varepsilon_{\rm{kin}} = 2 N_c\hspace{-1,5mm} \sum_{i=p,n}\hspace{-1.5mm} \left\{\int_0^\Lambda\hspace{-1.5mm} \frac{d^3p}{(2\pi)^3}\frac{p^2}{E_p^i}\left[1 - \theta(p_{F_i} - p)\right]\right\},
\end{equation}
$\rho_B=\rho_p+\rho_n$ and $\phi=\phi_p + \phi_n$ are the total baryonic and scalar densities obtained from
(\ref{rhobar}) and (\ref{phii}) respectively, with the number of
colors $N_c$ replaced by the spin degeneracy, and $\rho_3 = \rho_p - \rho_n$. 

The effective mass $M$ and the chemical potentials appearing in the thermodynamic potential $\Omega$ are also here determined by requiring that $\partial\Omega/\partial M = 0$ and $\partial\Omega/\partial p_{F_i} = 0$, resulting in
\begin{align}\label{Eq:Gap}
	M &= m - 2G_s\phi + 2G_{sv}\phi\rho_B^2 + 2 G_{s\rho}\phi\rho_3^2, \\
	\mu_i &= E_{p}^i + 2G_v\rho_B + 2G_{sv}\rho_B\phi^2 \pm 2G_\rho\rho_3 \nonumber +2G_{v\rho}(\rho_3^2\rho_B\pm \rho_B^2\rho_3) \pm 2 G_{s\rho}\rho_3\phi^2,
\end{align}
where  $'+'$ stands for $i=p$ and  $'-'$ stands for $i=n$,
and $E_{p}^i = \sqrt{M^2 + (p_{F_i})^2}$.
The equations of state can then be obtained using
(\ref{Exp_pressao_T}) and (\ref{Exp_energia_T}), taking
${M_{\rm{vac}}=m_N}$ representing the nucleon mass. More details
on the PPM NJL model and its parameterizations can be obtained in \cite{Pais2016}.

%%%%%%%%%%%%%%%%%%
\section{Binodals}
%%%%%%%%%%%%%%%%%%

\begin{figure}[!t]
\centering
\includegraphics{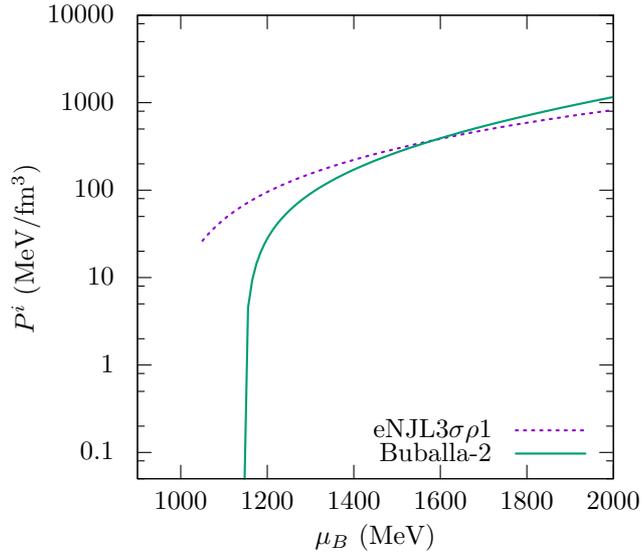}
\caption{Hadronic and quark matter pressures ($i$=H,Q) as functions of the
  baryonic chemical potential for symmetric matter.  \label{eos}}
\end{figure}

Before we start our discussion on the phase transition itself, we display in Figure \ref{eos}, the EoS of
both phases for two specific choices of parameters. The
discontinuities are related to the points where chiral symmetry is
restored and the points where the pressure becomes negative are
omitted. In Table  \ref{newtable} we show, for each of the quark and
the hadronic parameterizations used in the present work, the density
and chemical potential for  which chiral symmetry is restored.

The QCD phase-diagram is characterized by potentially multiple phases, whose phase separation boundaries are referred as \emph{binodals} \cite{Mueller1995}. Over those boundaries, the phases from the regions of either side of the boundary can coexist. The binodals may be determined using the Gibbs conditions \cite{Cavagnoli2011}:
\begin{align}
\mu_B^Q &= \mu_B^H , \label{Eq:ChemicalPotGibbsCondition}\\
T^Q &= T^H ,\\
P^Q &= P^H ,
\label{gibbs3}
\end{align}
where the indexes $H$ and $Q$ refer to the hadronic and quark phases. The chemical potentials are given by
\begin{align}
	\mu_B^H &= \frac{\mu_p + \mu_n}{2} , \\
	\mu_B^Q &= \frac{3}{2} (\mu_u + \mu_d) = 3 \mu_q.
\end{align}

\begin{table}[!t]
\centering
\caption{Values of $\rho_B$ and $\mu_B^i$ at the onset of chiral
  restoration for different parameterizations. $i$=H,Q.}
\begin{tabular}{ccc}
\toprule
Set & $\rho_B~(\rm{fm}^{-3})$ & $\mu_B^i~(\rm{MeV})$ \\
\midrule
Buballa-1 & 0.27 & 941 \\
Buballa-2 & 0.36 & 1035 \\
\midrule
PCP-0.0 & 0.29 & 1005 \\ 
PCP-0.1 & 0.24 & 1011 \\
PCP-0.2 & 0.20 & 1020 \\
PCP-0.3 & 0.17 & 1032 \\
PCP-0.4 & 0.17 & 1047 \\
PCP-0.5 & 0.17 & 1059 \\
\midrule
eNJL3$\sigma\rho 1$ & 1.0 & 1674 \\
eNJL2$m\sigma\rho 1$ & 1.0 & 1568 \\
\bottomrule
\end{tabular}
\label{newtable}
\end{table}

\begin{figure*}[!t]
\centering
\begin{minipage}[t]{.48\textwidth}
    \includegraphics[width=\linewidth]{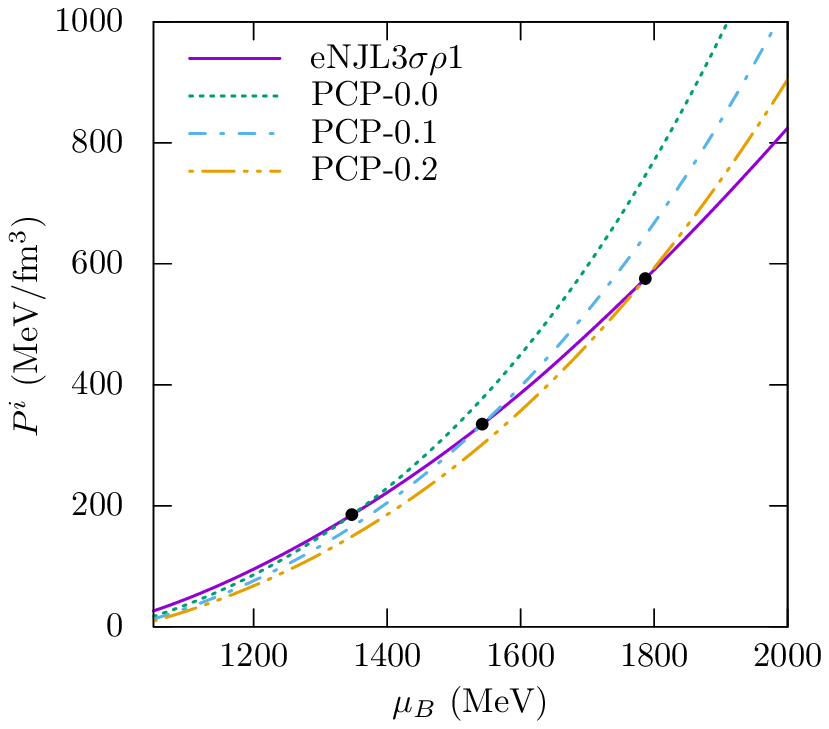}
    \caption{Examples of combinations of parameter sets for which
             hadron-quark phase transitions are allowed to happen in symmetric matter.\label{Fig:Intersection}}
\end{minipage}
\hfill
\begin{minipage}[t]{.48\textwidth}
	\includegraphics[width=\linewidth]{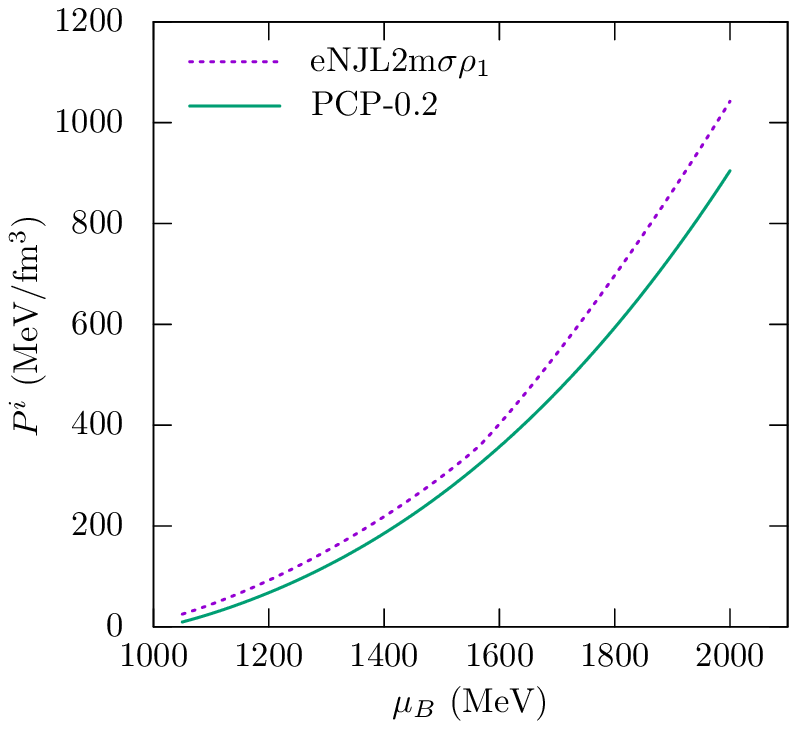}
	\caption{Example of a combination of parameter sets for which
             hadron-quark phase transitions are not allowed to happen in symmetric matter. \label{Fig:NoIntersection}}
\end{minipage}
\end{figure*}

At a certain fixed temperature ($T=0$ in the present context), the
phase coexistence condition may be obtained by plotting
${P^i\times\mu_B^i}$, $i = Q, H$, and looking for the intersection of
both curves. See Figure \ref{Fig:Intersection} for an example, where
the hadron pressure given by the eNJL3$\sigma\rho$1 parameterization
is plotted together with the quark pressure given by the PCP
parameterization of the NJL model for several choices of the vector
interaction strength $x$ such the coexistence of the hadron and the
quark phases occurs, allowing the phase transition to
happen. Otherwise, the absence of intersections imply that there are
no phase transitions allowed between the phases considered within a
specific pair of models, as shown in Figure \ref{Fig:NoIntersection},
where the hadronic matter is always
more stable. The existence of a hadron-quark phase transition
depends on both the quark matter and hadronic matter EoS:
the same  quark matter EoS, PCP-0.2, that  predicts a phase coexistence
with one hadronic EoS  (eNJL3$\sigma \rho$1) ceases to predict with a different one.

We determine the value of chemical potential $\mu_B^i$ for which the
phase transition takes place for all combinations of parameter
sets given in Tables~\ref{Tab:Parametros_NJL}  and
\ref{Tab:Parametros_eNJL}. At this point, we still restrict our
treatment of the hadron phase to symmetric matter due to the fact that
in our treatment of the quark phase the proportion of $u$ and $d$
quarks is always 50\% of each particle. This reflects the fact that
both particles are assumed to have the same bare masses and the same
chemical potentials. 
The results so obtained are displayed in
Table~\ref{Tab:Transition_chemical_pot}. In particular we may note
that no combinations involving the BuballaR-2 set with 
$G_v \neq 0$ give rise to a phase transition.  It should be pointed out the large
  differences among the chemical potential and density at the
  hadron-quark transition  predicted by the models
  considered.
Compatibility constraints between the
    hadronic and quark model should be
    imposed when describing the hadron-quark phase transition within a
  two-model description, which may reduce the phase transition
  uncertainties. In the present study, chiral
  symmetry is present in both the hadron and quark model.
Several compatibility constraints could be considered: i) the quark
phase should not be in a chiral broken  phase  at
deconfinement if the hadronic
phase is already in a chiral symmetric phase. This condition is
fulfilled for all cases discussed above. ii) a more restrictive
constraint would
be that at deconfinement the hadron and the quark phase have the same
chiral symmetry.  From
  Tables~\ref{newtable} and \ref{Tab:Transition_chemical_pot}, we may
  conclude that for  symmetric matter only quark models that predict a
deconfinement chemical potential above 1674  (1568) MeV are compatible
with eNJL3$\sigma\rho1$ (eNJL2m$\sigma\rho1$). i.e. Buballa-2,
Buballa-R2, and PCP-0.2;
iii)  however, we may also interpret that the deconfinement coincides with chiral symmetry restoration.
Moreover,  in fact the  eNJL2m$\sigma\rho1$ model has no chiral
symmetric phase because this is a model with a term breaking
explicitly the  chiral symmetry, and the chemical potential
indicated corresponds to half the vacuum mass.   In this scenario the
mixed phase between a pure hadronic and a pure quark matter phase
would be constituted by clusters of  non-chiral
symmetric hadronic matter in a background of chiral symmetric quark
matter, or the other way around;
iv) for asymmetric matter the possible scenarios are much more
complex because two or more conserved charges may be considered, and
the restoration of chiral symmetry will occur at different baryonic
densities or chemical potentials for different species. In the
following our discussion is based on interpretation iii) and we do
not discuss different scenarios corresponding to point iv).

Also constraints coming from experiments are needed to reduce
the uncertainty between models. One possibility is to use freeze-out information.

\begin{table}[!t]
\centering
\caption{Chemical potential, pressure, and
  barionic density at the coexistence point for different
  parameterization combinations in symmetric matter. $\rho_B$
  refers to the hadronic phase. For BuballaR-2, in this table, $G_v =
0$ and eNJL2m$\sigma\rho$1 presents no chiral symmetric phase (see the
text for details).}
\label{Tab:Transition_chemical_pot}
\begin{tabular}{ccccc}
\toprule
NJL SU(2) & Hadronic & $\mu_B$ (MeV) & $P$ ($\rm{MeV}/\rm{fm}^3$) & $\rho_B$ ($\rm{fm}^{-3}$) \\ 
\midrule
\multirow{2}{*}{Buballa-2} & eNJL2m$\sigma\rho$1 & 1674 & 504 & 1.420 \\
 & eNJL3$\sigma\rho$1 & 1567 & 356 &  0.812 \\
\midrule
\multirow{2}{*}{BuballaR-2} & eNJL2m$\sigma\rho$1 & 1729 & 586 & 1.497 \\
 & eNJL3$\sigma\rho$1 & 1585 & 373 & 0.839 \\
\midrule
\multirow{2}{*}{PCP-0.0} & eNJL2m$\sigma\rho$1 & 1312 & 158 & 0.506 \\
 & eNJL3$\sigma\rho$1 & 1348 & 185 & 0.553 \\
\midrule
PCP-0.1 & eNJL3$\sigma\rho$1 & 1544 & 336 & 0.780 \\
\midrule
PCP-0.2 & eNJL3$\sigma\rho$1 & 1787 & 576 & 1.088 \\
\bottomrule
\end{tabular}
\end{table}
\begin{figure*}[!t]
\centering
\begin{minipage}[t]{.48\textwidth}
    \includegraphics[width=\linewidth]{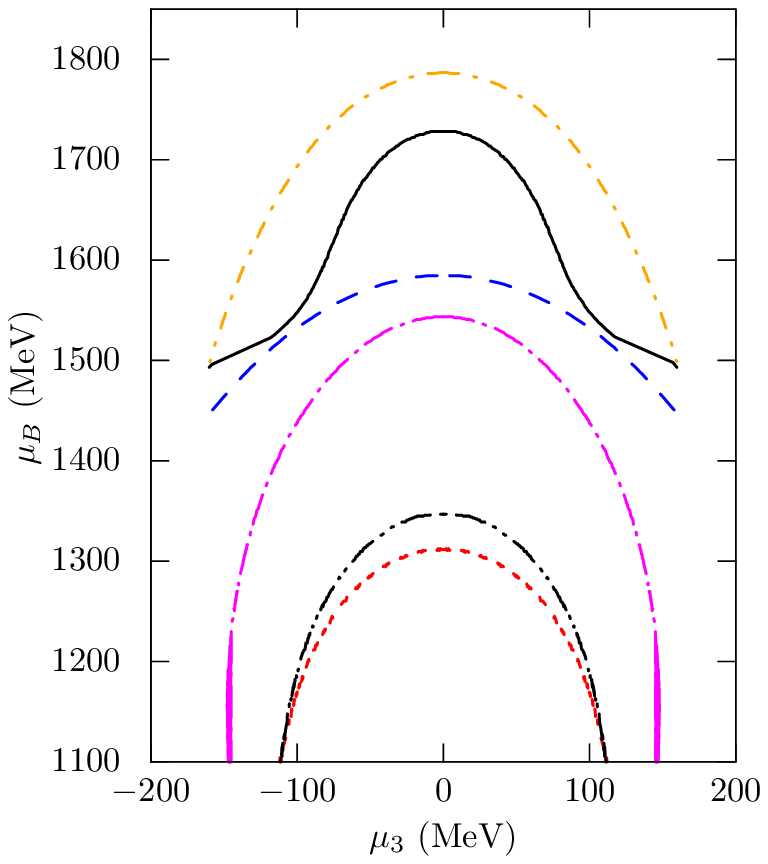}
    \caption{Baryonic chemical potentials at the coexistence point as a function of $\mu_3$ for: BuballaR-2 and eNJL2m$\sigma\rho$1 (black line), BuballaR-2 and eNJL3$\sigma\rho$1 (blue long-dashed line), PCP-0.0 and eNJL2m$\sigma\rho$1 (red short-dashed line), PCP-0.0 and eNJL3$\sigma\rho$1 (black double-dot dashed line), PCP-0.1 and eNJL3$\sigma\rho$1 (magenta long-dash dotted line), and PCP-0.2 and eNJL3$\sigma\rho$1 (orange short-dash dotted line).\label{Fig:binodals}}
\end{minipage}
\hfill
\begin{minipage}[t]{.48\textwidth}
    \includegraphics[width=\linewidth]{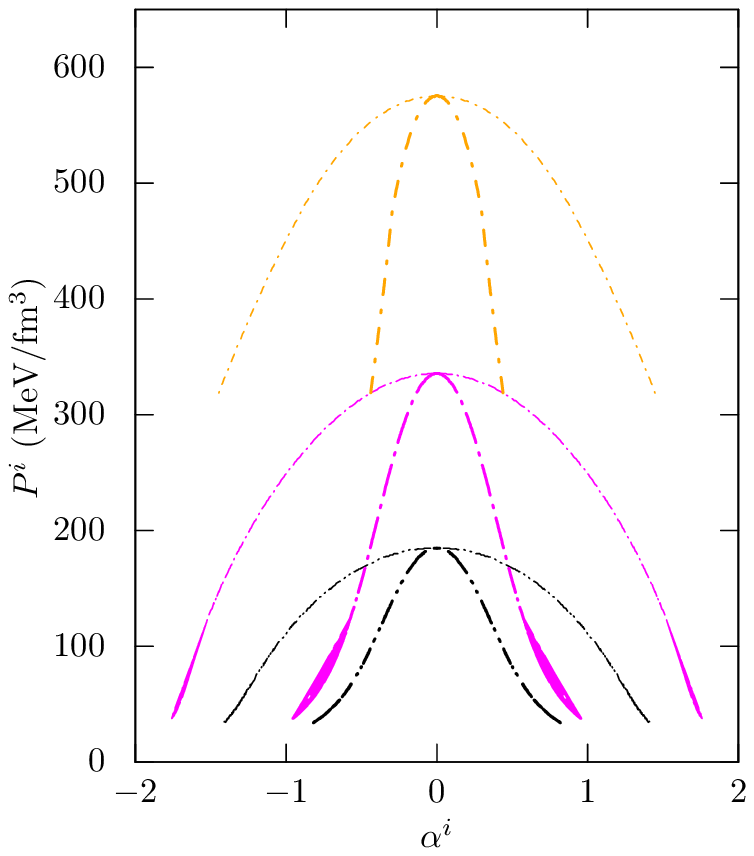}
    \caption{Pressure at the coexistence points for different
      asymmetry parameters using PCP-x for the quark phase and eNJL3$\sigma\rho_1$ for the hadron phase. The thick (internal) curve corresponds to the hadron phase, while the external one corresponds to the quark phase. The lines corresponds to: PCP-0.0 (black double-dot dashed), PCP-0.1 (magenta long-dash dotted line), and PCP-0.2 (orange short-dash dotted).\label{Fig:binodals_asymm}}
\end{minipage}
\end{figure*}
Up to this point, as stated previously, only symmetric matter was considered, since the
masses and chemical potentials of the $u$ and $d$ quarks in the
SU(2) NJL are identical. However, the conditions of phase coexistence 
are also important in asymmetric matter and to obtain the binodal
sections as a function of the system asymmetry, we use the
prescription given in \cite{muller1997}. The isospin chemical
potentials are defined as
\begin{align}
	\mu_3^H &= \mu_p - \mu_n , \\
	\mu_3^Q &= \mu_u - \mu_d,
\end{align}
and enforced to be identical according to the Gibbs conditions. 
The asymmetry parameters of the hadron and quark phases 
are respectively
\begin{equation}
\alpha^H=\frac{\rho_n-\rho_p}{\rho_n+\rho_p}, \qquad 
\alpha^Q= 3\frac{\rho_d-\rho_u}{\rho_d+\rho_u},
\end{equation}
in such a way that $0 \le \alpha^H \le 1$ (just nucleons) and  
$0 \le \alpha^Q \le 3$ (just quarks).

To obtain the \emph{binodals} we choose values of $\mu_B$ and $\mu_3$,
which determine the proton and neutron chemical potentials and,
through equation~\eqref{Eq:ChemicalPotGibbsCondition}, the chemical
potentials of the quarks. The $\mu_3$ parameter directly controls the
proton fraction of both phases. For each pair of values
$(\mu_B,\mu_3)$ we test the difference in pressure of both phases. If
this difference is smaller than a tolerance of $0.1~\textrm{MeV}$, we assume that there is a phase transition. This
procedure leads to the \emph{binodals} shown in
Figures~\ref{Fig:binodals} and~\ref{Fig:binodals_asymm}. 
The pressures shown for $\alpha = 0$ in Figure~\ref{Fig:binodals_asymm} correspond  to the intersections marked in Figure~\ref{Fig:Intersection}. 

From Table \ref{Tab:Transition_chemical_pot} and
Fig. \ref{Fig:binodals_asymm}, we can clearly see that the increase in
the value of $G_v$ causes a substantial modification on the transition
point, which reflects in the values of the pressure in the binodal
sections. 

As clearly stated in the Introduction, our aim is to obtain the
  QCD phase diagram with both hadronic and quark models based on the
  same underlying formalism, i.e., within different versions of the
  NJL model. The binodal section at zero temperature is the first
  step, but the inclusion of temperature and eventually, magnetic
  field will be performed. We next make a simple application of
  the phase transition to stellar matter to compare the results
  obtained with our formalism with the ones already existing in the literature. 

\section{Metastable stars}
%%%%%%%%%%%%%%%%%%%%%%%%%%

In order to describe compact star matter, leptons are included and
electric charge
neutrality and chemical equilibrium must be taken
enforced.
Leptons are
introduced in the system by adding them in the model Lagrangian density as a
free fermionic Lagrangian, i.e.,
\begin{equation}
\mathscr{L}=\bar{\psi}_l(i \gamma_\mu \partial^\mu -m_l)\psi_l,
\end{equation}
where $l$ refers to the leptons,  and unless stated otherwise, 
electrons and muons are considered, whose masses are, respectively, 0.511 MeV and 105.66 MeV.
Thus, the following constraints on chemical
potential and baryonic number density have to be imposed for hadronic star matter
\begin{align}
\mu_n&=\mu_p +\mu_e,\\
\rho_p &=\rho_e + \rho_{\mu},
\end{align}
 and, similarly, for quark star matter,
\begin{align}
\mu_s=\mu_d&=\mu_u+\mu_e, \\
\rho_e+\rho_{\mu}&=\frac{1}{3}(2\rho_u-\rho_d-\rho_s).
\end{align}
In both cases, $\mu_e=\mu_{\mu}$.

We next study the possibility of a hadron-quark phase transition to
take place in the interior of compact stars. Thus, we consider the EoS
obtained from the model presented in section \ref{NJLHM}, with
$\beta$-equilibrium and electric charge neutrality enforced, in the
description of hadronic stars. As for the quark matter, we consider the EoS derived in section
\ref{NJLSU3} to describe deconfined quark stars, also imposing 
$\beta$-equilibrium and electric charge neutrality. 
During the hadron-quark phase transition process, the composition of quark matter is not expected to
be $\beta$-stable \cite{2004ApJ...614..314B}. However, as we are mainly interested in the energetical
content of the final quark or hybrid star, this intermediate stage is
disregarded next.

\begin{figure}[!t]
\centering
\includegraphics{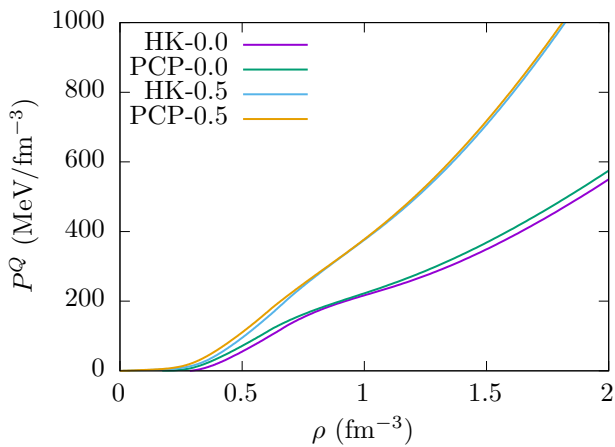}
	\caption{Pressure versus baryonic density for NJL SU(3) parameterizations
	with different values of x.  $\beta$-equilibrium and electric
        charge neutrality are enforced.	\label{fig:pcp-hk}}
\end{figure}

Figure \ref{fig:pcp-hk} shows the quark matter EoS for
some parameter choices, 
from where one can see the hardening effect of the vector
interaction in both situations, the same well known effect encountered in the SU(2) model
for hadronic matter without equilibrium conditions
\cite{njlv}.  The small bumps present in Figure
\ref{fig:pcp-hk}  are a characteristic of the chiral symmetry restoration associated
with the $s$ quarks.
Moreover, at large densities, after the total restoration of the
chiral symmetry, the densities of the three quarks are the same (1/3
of the total baryonic number density each).

In the same way as previously shown in Section III, we obtain  the
transition pressure and chemical potential 
which satisfy the Gibbs conditions
(\ref{Eq:ChemicalPotGibbsCondition}--\ref{gibbs3}), 
now enforcing $\beta$-equilibrium and charge neutrality within both phases. In stellar matter,
the baryonic and quark chemical potentials are usually defined in
terms of the EoS variables as 
\cite{2004ApJ...614..314B}
\begin{equation}
\mu_B^H =\frac{\varepsilon_H+P_H}{\rho}, \qquad
\mu_B^Q =\frac{\varepsilon_Q+P_Q}{\rho}, 
\end{equation}
taking $T=0$.
The results obtained for the coexistence points of hadron and quark
stellar matter are displayed in Table \ref{table:Transition_chemical_pot2}.
From it, we can see that the effect of the vector interaction on the phase transition is a
displacement of the phase transition point towards higher pressures
and higher chemical potentials.

\begin{table}[!t]
\caption{Chemical potential and pressure at the coexistence point for different
parameterization combinations for hadronic and three flavor quark
stellar matter with equilibrium conditions enforced.}\label{table:Transition_chemical_pot2}
\begin{center}
\begin{tabular}{cccc}
\toprule
 NJL SU(3) & Hadronic & $\mu_0$ (MeV) & $P_0$ (MeV/fm$^3$) \\
\midrule
 HK-0.0 & eNJL2m$\sigma\rho$1&    1399&196\\
 HK-0.1 & eNJL2m$\sigma\rho$1&   1529& 297\\
 HK-0.2 & eNJL2m$\sigma\rho$1&  1710&482\\
 HK-0.3 & eNJL2m$\sigma\rho$1&   2122 & 1144\\ 
\midrule
 HK-0.0 & eNJL3$\sigma\rho$1&      1349&154 \\
 HK-0.1 & eNJL3$\sigma\rho$1& 1462&227\\
 HK-0.2 & eNJL3$\sigma\rho$1&  1579& 313\\
 HK-0.3 & eNJL3$\sigma\rho$1 &  1709&422\\
 HK-0.4 & eNJL3$\sigma\rho$1 & 1863& 571\\ 
\midrule
 PCP-0.0 & eNJL2m$\sigma\rho$1&   1209&83\\
 PCP-0.1 & eNJL2m$\sigma\rho$1&   1420&211\\
 PCP-0.2 & eNJL2m$\sigma\rho$1&    1594&356\\
\midrule
 PCP-0.0 & eNJL3$\sigma\rho$1&  1170&64\\
  PCP-0.1 & eNJL3$\sigma\rho$1& 1328&143\\
  PCP-0.2 & eNJL3$\sigma\rho$1&  1481& 239\\
 PCP-0.3 & eNJL3$\sigma\rho$1&  1617& 344\\   
  PCP-0.4 & eNJL3$\sigma\rho$1&  1768&477\\
 PCP-0.5 & eNJL3$\sigma\rho$1&  1949&  663  \\
\bottomrule
\end{tabular}
\end{center}
\end{table}

Three diferent internal structures are next considered for the compact
star families: (i) hadronic stars modeled by the PPM NJL SU(2) 
equations of state; (ii) bare quark stars modeled by the NJL SU(3)
EoS; and (iii) hybrid stars, constituted by hadronic matter
in its outer region and  deconfined quark matter in the center. 
The equation of state for hybrid stars is
built from the hadronic and quark
EoS by performing a  Maxwell construction.
This method might seem naive since charge
neutrality is imposed only locally and results in the fact that 
the leptonic chemical potential suffers a discontinuity. But, as we
aim to study the macroscopic properties and the energetical content of
the compact stars, this construction suffices as shown in
\cite{mixed}. The BPS EoS \cite{bps} is also
included to the hadronic matter results to account for the description of the low-density matter in
the hadronic and hybrid stars outer crusts.

The family of possible compact stars are straightforwardly obtained by
using the equations of
state as input to the Tolman-Oppenheimer-Volkoff (TOV)
equations for the relativistic hydrostatic equilibrium \cite{ov,tov}. To solve the TOV
equations we need to impose boundary conditions given by $P(R)=0$ and $P(0)=P_c$,
where $R$ is the star radius and $P_c$ is the central pressure. 
In the following, $M(R)$ and $M_B(R)$ are the respectively the total gravitational mass and the total baryonic

At this point, a word of caution related to the bare quark stars is important. It is well known
  that the NJL SU(3) does not satisfy the Bodmer-Witten conjecture
  \cite{bubsqm}. However, the effects of a
   magnetic field not necessarily too strong and a small
  increase of temperature \cite {2013EPJC...73.2569D}  seem to be enough to guarantee that the
 quark matter acquires stability.

In the following, we investigate the conversion mechanism of
hadronic to hybrid stars. Similar analysis already exist in the
literature \cite{logoteta12, mixed, 2004ApJ...614..314B},
but models based on the same underlying field theory class in both hadron and quark
phases were never considered.

 If a compact star consisting only of hadrons and leptons in
  $\beta$-equilibrium, electrically neutral and with no fraction of
  deconfined quark matter, sustains a central pressure $P_C$ larger than
  the coexistence pressure of the hadron and quark phases, i.e. $P_0$,
  the hadronic star is said 
to be metastable to conversion to a  quark or hybrid star  
\cite{marquez17,2002NuPhS.113..268B, 2003ApJ...586.1250B, 2004ApJ...614..314B}. The possibility of the conversion depends on the values
of the hadronic star central pressure, $P_C$, and the  pressure that
satisfies the condition of phase coexistence, $P_0$, for a given pair
of EoS obtained from the respective models.

In Table \ref{table:ns-qs} we show some basic properties of
hadronic stars modeled with the parameterizations of the PPM NJL models
discussed, for stars with the maximum mass and for canonical stars with
$M=1.4~M_\odot$. Two of these results are of special relevance
following recent observational and theoretical advances, namely the
radius of the canonical neutron star ($R_{1.4M_\odot}$) and
the compactness of the maximum mass and the canonical star ($C_{M_{\text{max}}}$ 
and $C_{1.4M_\odot}$), defined as the ratio
between masses and radii of the respective compact stars. 
Both properties have been extensively discussed in the recent
literature \cite{Luiz2018, francesca}. Different hypotheses
lead to predictions of the radii of the canonical neutron star varying
from 9.7-13.9 km \cite{Hebeler2013} to 10.4-12.9 km \cite{Steiner2013} 
and from 10.1 to 11.1 km \cite{Ozel2016}. The results we show for the radii are not
compatible with the predictions of very small radii of
  \cite{Ozel2016} but lie within the other two constraints, as also obtained
in \cite{francesca} for a very large number of models.
Similarly, properties
of maximum mass configuration of quark and hybrid stars for some
parameter choices are shown in Table \ref{table:hys}. It is worth
  noticing that larger vector interaction parameters in the quark
  matter model result in more massive hybrid stars with smaller quark cores, reflecting
the stiffening of the EoS discussed in \cite{njlv}.
Indeed,  following the
effect of the vector interaction in the displacement of the phase
transition point to higher pressures, as $P_0$ approaches  the maximum $P_C$ of the metastable star family, the deconfined quark matter core is possible only inside the  most massive stars.  As a result, the TOV stable solutions for hadronic and hybrid EoS  differ only  for a narrow set of stars where the condition $P_C\geq P_0$ is  fulfilled.
The compactness of both pure hadronic and hybrid canonical star
are close to the one recently measured for an isolated neutron star
\cite{comp} as being equal to $0.105 \pm 0.002$.

 \begin{table}[!t]
\caption{Stellar macroscopic properties obtained with the two PPM eNJL parameterizations. The first set of values refers to the maximum mass star and the later to the canonical star.}\label{table:ns-qs}
\begin{center}
\begin{tabular}{lcc}
\toprule
& eNJL2m$\sigma\rho$1 & eNJL3$\sigma\rho$1 \\
\midrule
$M_{\text{max}}$ ($M_\odot$) & 2.02 & 2.19 \\
${M_B}$ ($M_\odot$) & 2.33 & 2.56 \\
$R$ (km) & 11.19 & 11.37\\
$C_{M_{\text{max}}}$ ($M_\odot$/km)& 0.180&0.192\\
$\rho_C$ (fm$^{-3}$)& 0.981 &0.966 \\
$\mu_C$ (MeV)& 1623 &1781\\
$P_C$ (MeV/fm$^3$) & 363 & 489\\
\midrule
$R_{1.4M_\odot}$ (km) & 12.20&12.94\\
$C_{1.4M_\odot}$ (km/$M_\odot$)& 0.114&0.108\\
\bottomrule
\end{tabular}
\end{center}
\end{table}

 \begin{table*}[!t]
\caption{Stellar macroscopic properties of quark and hybrid stars, obtained with some diferent EoS parameterizations for the phases. 
The first set of values refers to the maximum mass star and the second to the canonical star.
For the hybrid stars, $\rho_{H}$ and $\rho_{Q}$  denote the densities of the metastable and quark matter at the phase coexistence point,
and $M_{H\text{-}Q}$ denotes the gravitational mass of the less massive star that sustains a deconfined quark core. The units are the same as the Table \ref{table:ns-qs}. \label{table:hys}}
\label{table:qs}
\begin{center}
%\begin{tabular}{ccccccccc|cc|ccc}
%\toprule
%\multicolumn{2}{c}{Equation of State}& $M_{\text{max}}$ & ${M_B}$ &$R$&$C_{M_{\text{max}}}$& $\rho_C$ & $\mu_C$ & $P_C$&$R_{1.4M_\odot}$ &$C_{1.4M_\odot}$ 
%                                     & $\rho_{H}$ & $\rho_{Q}$&$M_{H\text{-}Q}$\\
%\midrule
%\parbox[t]{2mm}{\multirow{3}{*}{\rotatebox[origin=c]{90}{Quark}}}  & {PCP-0.0}   & 1.63  & 1.81 &9.90&0.164  & 1.035 & 1408  & 230 &10.00&0.140     \\ 
%                             & {PCP-0.2}   & 1.79  & 1.97  &10.19&0.175   & 0.995  & 1527  & 283&10.43&0.134       \\ 
%                             & {PCP-0.5}   & 1.97& 2.15 &10.79&0.182 & 0.915 & 1667 & 332 &11.05&0.126               \\
%\\
%\parbox[t]{2mm}{\multirow{6}{*}{\rotatebox[origin=c]{90}{Hybrid}}} & eNJL2m$\sigma\rho$1, PCP-0.0& 1.80&2.03&11.60&0.155& 0.910&1380&202&12.20&0.114&0.487&  0.527& 1.62          \\
%                             & eNJL2m$\sigma\rho$1, PCP-0.2& 2.02&2.33& 11.23&0.179&1.084&1594&356 &12.20&0.114&  0.979&1.185 &--             \\
%                             & eNJL3$\sigma\rho$1, PCP-0.0 & 1.63 & 1.81&12.02&0.135 & 1.021 & 1408& 227 &12.94&0.108&0.421&0.477 & 1.57          \\
%            & eNJL3$\sigma\rho$1, PCP-0.1 & 1.97 & 2.25&12.25&0.160 & 0.834 & 1408& 205 &12.94&0.108&0.564&0.648 & 1.93          \\
%                             & eNJL3$\sigma\rho$1, PCP-0.2 & 2.18 & 2.55&12.13&0.179 & 0.820 & 1481& 239 &12.94&0.108&0.700&0.873&--  \\
%                             & eNJL3$\sigma\rho$1, PCP-0.4 & 2.19 & 2.57&11.39&0.192& 1.118& 1768 & 477&12.94&0.108&0.955&1.234 &-- \\
%\toprule
%\end{tabular}

\begin{tabular}{clcccccccccccc}
\toprule
&&& \multicolumn{3}{c}{Quark Star} && \multicolumn{7}{c}{Hybrid Star} \\
\cmidrule{4-6} \cmidrule{8-14}
&&&&&&& \multicolumn{2}{c}{eNJL2m$\sigma\rho$1} && \multicolumn{4}{c}{eNJL3$\sigma\rho$1} \\
\cmidrule{8-9}\cmidrule{11-14}
& PCP-$x$:                                  && 0.0     & 0.2    & 0.5    && 0.0   & 0.2   && 0.0    & 0.1   & 0.2    & 0.4 \\
\cmidrule{2-2} \cmidrule{4-6} \cmidrule{8-9} \cmidrule{11-14}
\parbox[t]{2mm}{\multirow{15}{*}{\rotatebox[origin=c]{90}{Properties}}} & \multicolumn{1}{|l}{$M_{\text{max}}$} 
                                            && 1.63    & 1.79	& 1.97	 && 1.80   & 2.02  && 1.63  & 1.97  & 2.18   & 2.19 \\
& \multicolumn{1}{|l}{${M_B}$}	            && 1.81    & 1.97	& 2.15	 && 2.03   & 2.33  && 1.81  & 2.25  & 2.55   & 2.57 \\
& \multicolumn{1}{|l}{$R$}	            && 9.90    & 10.19	& 10.79	 && 11.60  & 11.23 && 12.02 & 12.25 & 12.13  & 11.39 \\
& \multicolumn{1}{|l}{$C_{M_{\text{max}}}$} && 0.164   & 0.175 	& 0.182	 && 0.155  & 0.179 && 0.135 & 0.160 & 0.179  & 0.192 \\
& \multicolumn{1}{|l}{} & \\
& \multicolumn{1}{|l}{$\rho_C$}	            && 1.035   & 0.995	& 0.915	 && 0.910  & 1.084 && 1.021 & 0.834 & 0.820  & 1.118 \\
& \multicolumn{1}{|l}{$\mu_C$}              && 1408    & 1527	& 1667	 && 1380   & 1594  && 1408  & 1408  & 1481   & 1768 \\
& \multicolumn{1}{|l}{$P_C$}                && 230     & 283	& 332	 && 202	   & 356   && 227   & 205   & 239    & 477 \\
& \multicolumn{1}{|l}{$R_{1.4M_\odot}$}     && 10.00   & 10.43	& 11.05	 && 12.20  & 12.20 && 12.94 & 12.94 & 12.94  & 12.94 \\
& \multicolumn{1}{|l}{$C_{1.4M_\odot}$}     && 0.140   & 0.134	& 0.126	 && 0.114  & 0.114 && 0.108 & 0.108 & 0.108  & 0.108 \\
& \multicolumn{1}{|l}{} & \\
& \multicolumn{1}{|l}{$\rho_{H}$}           && 	       &	&	 && 0.487  & 0.979 && 0.421 & 0.564 & 0.700  & 0.955 \\
& \multicolumn{1}{|l}{$\rho_{Q}$}           && 	       &	&	 && 0.527  & 1.185 && 0.477 & 0.648 & 0.873  & 1.234 \\
& \multicolumn{1}{|l}{$M_{H\text{-}Q}$}     &&	       &	&	 && 1.62   & --	   && 1.57  & 1.93  & --     & -- \\
\bottomrule
\end{tabular}
\end{center}
\end{table*}

\begin{figure}[!t]
\centering 
\includegraphics{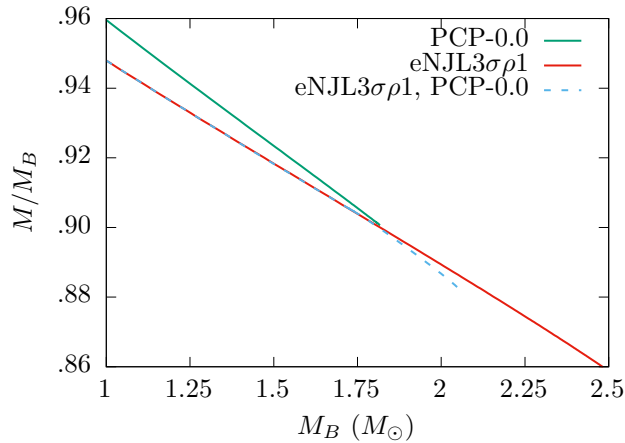}
	\caption{Ratio between the gravitational and baryonic masses versus baryonic mass of hadronic, hybrid and quark stars, for diferent EoS parameterizations.} \label{masses2}
\end{figure}

Moreover,  we can see that the central pressures $P_C$ of 
the hadronic stars are larger than some of the coexistence pressure values
$P_0$, as shown in Table \ref{table:Transition_chemical_pot2},
notably for smaller values of the 
vector interaction parameter x in the quark matter modeling.
This is the first condition that enables
the conversion of a metastable neutron star 
into a quark or hybrid star. The other condition is that the
gravitational mass of the initial metastable hadronic
 star must be bigger than the gravitational mass of the final star, either quark or hybrid star, for
 a given baryonic mass, so that the conversion can be exothermal in
 rest while respecting the baryonic number conservation
 \cite{marquez17}. In Figure \ref{masses2} we ilustrate the results by
 plotting the ratio between the gravitational and baryonic masses with
 respect to the baryonic mass, in a way that highlights the small differences between the curves while preserving the interpretation that the conversion is energetically allowed only if the final configuration is below the initial one for $M_B$ fixed.

 The gravitational masses of quark stars are 
bigger than the gravitational mass of the hadronic star with the same
baryonic mass, which is already expected from previous results in
literature, e.g., \cite{bubsqm}. Follows that the conversion of a 
hadronic star to a bare quark star is always energetically forbidden for the parameterizations considered in this work, even in
cases where it would be allowed by the Gibbs thermodynamic condition. 
This feature can be better understood looking for some notable cases.
For  the {PCP-0.5} case, which results in a quark star with $P_C=332$ MeV/fm$^3$,  we 
see in Table \ref{table:Transition_chemical_pot2} that the conversion is allowed by the Gibbs criteria only if
the  eNJL3$\sigma\rho$1 hadronic matter is used in the modeling of the metastable star.
However, the coexistence pressure is much higher than the ones
sustained by the compact stars described by each phase.  This feature prevents the conversion
to take place, since it would occur at constant $P$, i.e., both
initial and final should sustain $P_C\geq P_0$.
Taking the PCP-0.2 case, instead, the quark star sustains $P_C=283$
MeV/fm$^3$, which allows a hadron-quark
coexistence point with both hadronic matter parameterizations, as
seen in Table \ref{table:Transition_chemical_pot2}. If the
eNJL2m$\sigma\rho$1 hadronic matter is considered, we have
 $P_C^H=363>P_0=356>P_C^Q=283$ MeV/fm$^3$. It means that, despite the 
metastable hadronic star
bulk is overpressured enough to allow the phase transition to the
PCP-0.2 quark matter, there are not such final compact object 
constituted by the latter phase.  The metastable star decays into
  a black hole. In other words, this set
fulfills the thermodynamic criteria but the astrophysical conditions
do not allow the formation of a stable quark star.
The last set to be analyzed is when the PCP-0.2 quark matter is
compared with the eNJL3$\sigma\rho$1 hadronic matter. 
In this case $P_C^H=489>P_0=239$ MeV/fm$^3$ and $P_C^Q=283>P_0=239$ MeV/fm$^3$.
The main imposition to the hadron-quark phase transition to take place
inside metastable compact stars is to have $P_C\geq P_0$ for both
stars, which is fulfilled by this choice of models. Nevertheless,
a  conversion process that preserves baryonic mass, requires that the final state
has the same baryonic mass and a smaller gravitational mass. Since 
the quark star has a larger  gravitational mass, the conversion is
forbidden due to energy arguments.  

A different situation occurs when hybrid stars are considered. In
Figure \ref{masses2}, we can see the hybrid star
family curve differs from the respective pure hadronic star family  for stars with a
central density above $P_0$, i.e., for hadronic metastable stars
massive enough to sustain the conversion of their core from the
hadronic matter to a deconfined quark matter bulk. It follows from
previous results that the branches where the conversion is allowed are
bigger for smaller values of the vector interaction parameter $x$ in
the quark matter modeling. In fact, we get a quark core only for a low
enough $x$ value, which is 0.12 for nuclear matter model 
eNJL3$\sigma\rho1$ \cite{Pais2016} and 0.1  for eNJL2m$\sigma\rho1$,
as can be seen from table \ref{table:qs} by comparing the values of
$\rho_C$ with the values of $\rho_Q$. Stable stars are only possible
if $\rho_C$ is larger than $\rho_Q$ and $P_0$ larger than $P_C$. Again, analysing
the results shown in tables \ref{table:Transition_chemical_pot2} and
\ref{table:qs}, we see that in most cases these pressures are identical and a
stable star with a quark core is not sustainable.
Another feature worth noticing is that, even when the conversion from a hadronic to a hybrid star is allowed, the mass-energy difference of the initial and final objects are always small (a narrow gap of the order of $10^{-3}$--$10^{-2}$ $M_\odot$).

%%%%%%%%%%%%%%%%%%%%%
\section{Conclusions}
%%%%%%%%%%%%%%%%%%%%%

In this work we have revisited the study of hadron-quark phase transition at zero
temperature with different extensions of the NJL model, which are more
appropriate to describe systems where chiral symmetry is an important
ingredient. 

We have first analyzed possible phase transitions from 
a hadron phase described by the PPM NJL with nucleons to a quark phase
described by  the SU(2) NJL with the inclusion of a vector interaction
whose strength is arbitrary. We have considered symmetric matter and
verified that the  existence of a hadron-quark phase transition
depends on both the quark matter and hadronic matter EoS. For a given
hadronic EoS there is a limiting $G_V$ value above which the
transition ceases to exist. For a given quark EoS the deconfinement 
transition does not occur if the hadronic EoS is too soft. We have
considered two models that reproduce well accepted properties of
nuclear matter at saturation, as well as experimental results obtained
from collective flow data in heavy-ion collisions \cite{danielewicz02}
and from the KaoS experiment \cite{kaos}. The  existing high density constraints
are, however, too much model dependent and it is not clear how
reliable they are.
Another manifestation of the dependence of the results on the choice
of parameters is the range of chemical potentials for which the
transition takes place: it spans from 1312~MeV to 1787~MeV, a
  30\% difference in relation to the lowest value. In terms of the transition
densities, this translates into a density that can be as low as 0.5
fm$^{-3}$ or as high as  1.5
fm$^{-3}$, with one of the quark  models predicting much smaller
transition densities.  This indicates a strong
parameter dependence  and experimental constraints are needed.
Chiral symmetry may impose some constraints. In our study
  deconfinement and chiral symmetry restoration are coincident
  transitions, but different scenarios could occur, as the 
  chiral symmetry restoration before the
  deconfinement transition, corresponding to a quarkyonic phase,  and this scenario would certainly impose
  strong constraints on the hadronic EoS.
We have next analyzed asymmetric systems
and, whenever possible, binodal sections were obtained. Both pressure
and chemical potential increase drastically with the increase of the
vector interaction strength in the quark sector. For asymmetric
  matter we may expect the appearance of quark matter at smaller
  densities according to \cite{Cavagnoli2011}. 
As a next step on this analysis, we plan to expand our results to include
finite temperature in the system and obtain the complete binodal
sections.

Afterwards, for a more complete treatment, we have investigated 
phase transitions from hadronic stellar matter (same PPM NJL model,
but subject to $\beta$- equilibrium and charge neutrality) to quark
stellar matter. In this case,  flavor conservation is not possible
because the hadronic phase only includes nucleons and the quark phase
should contain strangeness to satisfy (although barely) the
Bodmer-Witten hypothesis.  We have seen
that, in general, the phase transition pressure and chemical
potentials increase with the increase of the vector interaction
strength in the quark sector, as before. Then, we have seen that the conversion of
metastable stars to quark stars within the studied models is virtually impossible if the
condition that the gravitational mass of the hadronic star has to be
larger than the gravitational mass of the quark star at the same
baryonic mass is imposed. Nevertheless, the conversion from
  hadronic to hybrid stars is possible, but the mass-energy difference
  between both objects is very small.

%The column width is: \the\columnwidth

\acknowledgments
DPM and KDM acknowledge partial support from CNPq (Brazil). This work is also a 
part of the project CNPq-INCT-FNA Proc. No. 464898/2014-5.
C.A.G. is very thankful to DPM for the hospitality and supervision
during his post-doctoral 
internship at UFSC.
C.A.G. and M.D.A. acknowledge the support of Funda\c{c}\~ao de Amparo \`a Pesquisa e Inova\c{c}\~ao do Estado de Santa Catarina (FAPESC) under Grant No. 2017TR1761.
C.P. acknowledges financial support by Fundação para a Ciência e
Tecnologia (FCT) Portugal under the projects No. UID/FIS/04564/2016 and POCI-01-0145-FEDER-029912 
(with financial support from POCI, in its FEDER component, and by the FCT/MCTES budget through 
national funds (OE)).

%%%

%\appendix
%\section{Appendix title}

%%%

\bibliographystyle{JHEP} 
\bibliography{references}

%%%%%%%%%%%%%%
%%%%%%%%%%%%%%
%%%%%%%%%%%%%%
\end{document}